\documentclass[11pt]{article}
\newcommand{\be}{\begin{eqnarray}}\newcommand{\beq}{\begin{equation}}
\newcommand{\ee}{\end{eqnarray}}\newcommand{\eeq}{\end{equation}}
\newcommand{\ep}{\epsilon}\newcommand{\eps}{\varepsilon}
\newcommand{\la}{\lambda}\newcommand{\De}{\Delta}

\usepackage{graphicx} 
\usepackage{amssymb}
\usepackage{graphicx,amssymb,amsmath} 
\title{
On the Fokker-Planck
approximation in the kinetic equation of multicomponent classical nucleation theory
}
\author{Yuri S. Djikaev,\thanks{Corresponding author. 
E-mail: idjikaev@buffalo.edu}
\hspace{0.1cm} \hspace{0.1cm}Eli Ruckenstein,\thanks{deceased}\hspace{0.1cm} and 
\hspace{0.1cm} 
Mark Swihart\thanks{E-mail: swihart@buffalo.edu}
\\ Department of Chemical and Biological  Engineering, SUNY at Buffalo, \\ Buffalo, New York  14260 
}
\date{}
\renewcommand{\baselinestretch}{2}
\begin{document}
\renewcommand{\baselinestretch}{1}
\maketitle
\begin{abstract}

We examine the validity of the Fokker-Planck equation with linear force coefficients as an
approximation to the kinetic equation of nucleation in homogeneous isothermal multicomponent condensation.
Starting with a discrete equation of balance governing the temporal evolution of the distribution function of
an ensemble of multicomponent droplets and reducing it (by means of Taylor series expansions) to the
differential form in the vicinity of the saddle point of the free energy surface,  we have identified the
parameters whereof the smallness is necessary for the resulting kinetic equation to have the form of the
Fokker-Planck equation with linear (in droplet variables) force coefficients. The ``non-smallness" of these
parameters results either in the appearance of the third or higher order partial derivatives of the
distribution function in the kinetic equation or in its force coefficients becoming non-linear functions of
droplet variables, or both; this would render the  conventional kinetic equation of multicomponent nucleation
and its predictions inaccurate. As a numerical illustration, we carried out calculations for isothermal
condensation in five binary systems of various  non-ideality at $T=293.15$ K:  1-butanol--1-hexanol,
water--methanol, water--ethanol, water--1-propanol, water--1-butanol. Our results suggest that under typical
experimental conditions the  kinetic equation of binary nucleation of classical nucleation theory may require
a two-fold modification and,  hence, the conventional expression for the steady-state
binary nucleation rate may not be adequate for the  consistent comparison of theoretical predictions with
experimental data.

\end{abstract}
\renewcommand{\baselinestretch}{2}
\newpage

\section{Introduction}

Nucleation is the initial stage of any homogeneous 
first order phase transition$^{1-3}$ that does not occur 
as spinodal decomposition. 
At the nucleation stage of condensation, hereinafter the sole subject of consideration for the sake of concreteness, 
the  initial growth of nascent particles (droplets) of the liquid phase is due exclusively to fluctuations; 
the association  of two molecules and the subsequent association of the third, fourth, and additional molecules is
thermodynamically unfavorable (i.e., is accompanied by an increase in the free energy of the system),   but
does occur owing to fluctuations.  However, after a droplet attains some critical size (and composition, in the 
case of multicomponent condensation), the  incorporation of each additional molecule 
becomes   thermodynamically favorable (i.e., is accompanied by a decrease in the free energy of the system), 
and the droplet grows irreversibly.  The free energy of formation of the critical droplet (often
referred to as a ``nucleus") determines the height of the activation, or nucleation, barrier. 

The distribution function of an ensemble of droplets with respect to the independent variables of state of a
droplet represents the main object of interest in any theory of homogeneous condensation. In particular, such
a distribution of near-critical droplets determines the nucleation rate. 
The temporal evolution of the distribution of near-critical droplets
is governed by the equation whereof the finite-differences form is often referred to as a ``balance equation" 
whereas its differential form is called a ``kinetic equation" of nucleation. 

In the case of isothermal nucleation (in which the temperature of any single droplet is constant and  equal to
the temperature of the surrounding vapor-gas medium), the kinetic equation of nucleation is assumed to be well
approximated by the Fokker-Planck equation. In the case of non-isothermal nucleation, where the possibility of
the deviation of the droplet temperature from that of the surrounding medium 
is taken into account, the Fokker-Planck approximation has been shown to be inadequate to describe  the
evolution of the distribution function with respect to the droplet temperature. Hereinafter, we do not
consider the nonisothermal case, but focus on the applicability of the Fokker-Planck approximation to the
kinetic equation of isothermal multicomponent nucleation. 

\section{The Fokker-Planck approximation in the kinetic equation of homogeneous isothermal nucleation}

In the kinetic theory of homogeneous isothermal condensation, the equation governing the temporal evolution of
the distribution of near-critical droplets with respect to the number of molecules in a droplet (or with
respect to numbers of molecules of different components in a droplet) is conventionally considered to have the
Fokker-Planck form. The  accuracy of such an assumption for unary nucleation has been thoroughly examined by
Kuni and Grinin.$^{4}$ On the other hand, its accuracy in the case of multicomponent nucleation has been
studied very little. We are aware of only two relevant papers; one by Kuni {\it et al.},$^{5}$ who
qualitatively outlined the general principles of the Fokker-Planck approximation in a kinetic equation of
nucleation, and the other by Kurasov,$^{6}$ who qualitatively discussed this issue in the case of
nonisothermal binary  nucleation. In this section, we will first briefly outline the results of Kuni and
co-workers concerning this issue in unary nucleation (subsection 2.1) and then attempt to shed some
light on the validity of the Fokker-Planck approximation in the kinetic equation of homogeneous isothermal
multicomponent nucleation (subsection 2.2). 

\subsection{Unary nucleation}

Consider an ensemble of one-component droplets within the metastable vapor (of the same component) at 
temperature $T$, and denote the number of molecules in a droplet by $\nu$; this will be the only variable of
state if nucleation is isothermal (i.e., the droplet temperature is constant and equal to $T$).  The
capillarity approximation,$^{7}$ whereon the thermodynamics of classical nucleation theory (CNT) is based,
requires the  liquid droplets to be sufficiently large, with $\nu\gg 1$, and of spherical shape, with sharp
boundaries, and uniform density inside. 
The 
metastability of the one-component vapor is usually characterized by the saturation ratio
$\zeta=n/n_{1\infty}$, where $n$ is the number density of vapor molecules  and $n_{1\infty}$ is the 
number density of molecules of the vapor in equilibrium with its bulk liquid at the system  
temperature. Clearly, the vapor-to-liquid transition can occur only if $\zeta>1$; at sufficiently 
large values of $\zeta$, it will occur as spinodal decomposition, otherwise it will proceed via nucleation. 

Denote the distribution function of droplets with respect $\nu$ at time $t$ by $g(\nu,t)$. Assuming that the
droplets exchange matter with the vapor via the absorption and emission of single molecules, the temporal
evolution of $g(\nu,t)$ is governed by the balance equation
\beq 
\frac{\partial g(\nu,t)}{\partial t}=-
\left[(W^+(\nu)g(\nu,t)-W^-(\nu+1)g(\nu+1,t))-
(W^+(\nu-1)g(\nu-1,t)-W^-(\nu)g(\nu,t))\right],
\eeq
where $W^+(\nu) and W^-(\nu)$ are the numbers of molecules that a droplet $\nu$ absorbs and emits,
respectively, per unit time. A differential equation governing the temporal evolution of $g(\nu,t)$ can be
obtained from the discrete  balance equation (1) through  the Taylor  series expansions of  $W^-(\nu\pm
1),W^+(\nu\pm 1)$, and $g(\nu\pm 1, t)$ (on its RHS) with respect to the deviation of their arguments from 
$\nu$. 

According to classical thermodynamics, the equilibrium distribution function has the form
\beq 
g_e(\nu)=n\exp[-F(\nu)], 
\eeq 
where $F(\nu)$ is the free energy of formation of a droplet of size $\nu$ (in units $k_BT$, $k_B$ being the
Boltzmann constant). In the framework of CNT, $F(\nu)$ can be written$^{4,5}$ as 
\beq 
F(\nu)=-b\nu+a\nu^{2/3},
\eeq
where $b=\ln\zeta$ and $a=4\pi(3v_l/4\pi)^{2/3}(\sigma/k_BT)$, with $v_l$ being the volume per molecule in the
liquid phase and $\sigma$ the droplet surface tension (assumed to be equal to the surface tension of the bulk
liquid).  When condensation occurs via nucleation, 
the function $F(\nu)$ has a maximum at some  $\nu_c=(2a/3b)^3$. A droplet with $\nu=\nu_c$ is called
``nucleus"; the subscript ``c" will mark quantities related thereto. 

Defining the quantity $\De\nu_c$ by the equality  
\beq 
\frac1{2}|F_c''|(\De\nu_c)^2=1,
\eeq
where $F''=\partial^2 F/\partial \nu^2$, Kuni and Grinin$^{4}$ pointed out that the free energy of droplet
formation $F(\nu)$ and   equilibrium distribution $g_e(\nu)$ can be accurately represented as 
\beq
F(\nu)\simeq F_c+\frac1{2}F_c''(\nu-\nu_c)^2 ,\;\;\;
g_e(\nu)\simeq g_e(\nu_c)\exp[-\frac1{2}F_c''(\nu-\nu_c)^2], 
\eeq
respectively, over the entire region $(|\nu-\nu_c|\lesssim\De\nu_c)$ of the substantial change of $g_e(\nu)$ 
in the vicinity of $\nu_c$ if 
\beq 
\De\nu_c/\nu_c\ll 1. 
\eeq
The relative inaccuracy of representations (5)  within the near-critical region $|\nu-\nu_c|\lesssim\De\nu_c$
is of the order of $\De\nu_c/\nu_c$. 

As clear from eq.(5), $\De\nu_c$ represents the characteristic scale of the substantial change of the
equilibrium distribution function $g_e(\nu)$ in the vicinity of $\nu_c$. Moreover, Kuni and Grinin$^{4}$ 
showed that in that vicinity $\De\nu_c$ also represents the characteristic scale of the substantial  change of
the steady-state distribution function $g_s(\nu)$ as well as of the distribution $g(\nu,t)$, so  
\beq
\frac1{g(\nu,t)}\frac{\partial g(\nu,t)}{\partial \nu} \sim 
\frac1{g_s(\nu)}\frac{d g_s(\nu)}{d\nu} \sim \frac1{g_e(\nu)}\frac{d g_e(\nu)}{d\nu} 
\sim \frac1{\De\nu_c}.
\eeq 

The absorption rate $W^+(\nu)$ of a droplet (in eq.(1)) is determined from the gas-kinetic theory,$^{1-5}$
\beq
W^+(\nu)=\frac1{4}n\;\overline{v}_TA(\nu),
\eeq
where $\overline{v}_T=\sqrt{8k_BT/\pi m}$  is the mean thermal velocity  of vapor molecules (of mass $m$) and
$A(\nu)=4\pi(3v_l/4\pi)^{2/3}\nu^{2/3}$ is the  surface area of the droplet. On the other hand, the
droplet emission rate $W^-(\nu)$ is determined through $W^+(\nu)$ from the principle of detailed balance,
stipulating that for the equilibrium distribution of droplets 
$W^-(\nu)g_e(\nu)= W^+(\nu-1)g_e(\nu-1)$, 
so that, according to eq.(2),
\beq
W^-(\nu)= W^+(\nu-1)\exp[F(\nu)-F(\nu-1)].
\eeq

Carrying out the Taylor series expansions on the RHS of eq.(1) and taking into account eqs.(3)-(9), Kuni and
Grinin$^{4}$ showed that for the resulting differential equation  in the near-critical region
$|\nu-\nu_c|\lesssim\De\nu_c$ to be accurately approximated by the Fokker-Planck equation 
\beq
\frac{\partial g(\nu,t)}{\partial t}=-W^+_c\frac{\partial}{\partial \nu}\left(-F'(\nu)-\frac{\partial}
{\partial \nu}\right)g(\nu,t)  
\eeq
with the drift/force coefficient $F'(\nu)=\partial F/\partial \nu$ a linear function of $\nu$, the strong
inequality 
\beq
\frac1{\De\nu_c}\ll 1
\eeq
must be fulfilled {\em in addition} to condition (6). The parameters 
$\De\nu_c/\nu_c$ and  $1/\De\nu_c$ can be considered to represent the {\em small 
parameters} of the macroscopic theory of condensation.

Thus, for the Fokker-Planck approximation to be suitable in the kinetic equation of
nucleation, there must exist some near-critical region whereof the half-width $\De\nu_c$, defined by
constraint (3), satisfies the following requirements:\\
a) $\De\nu_c$ is large enough to represent the  characteristic scale of substantial change of the
equilibrium distribution function in the vicinity of $\nu_c$.\\  
b) $\De\nu_c$ is small enough that the quadratic approximation (eq.(5)) for the free  energy of formation
is acceptable across the entire near-critical vicinity.\\
c) $\De\nu_c$ is much greater than the elementary change of the droplet variable; this requirement ensures
that in the Taylor series expansions of the RHS of eq.(1) the terms with the third and higher order
derivatives of the distribution function $g(\nu,t)$  can be neglected compared to the term containing the
second order derivative of $g(\nu,t)$.\\
Note that ({\em in unary condensation theory only!})  the requirements b) and c) are expressed through strong
inequalities (6) and (11), whereas   the requirement a), expressed as the operator estimates in eq.(7), is
automatically satisfied due to constraint (4) if the requirement b) is satisfied.  

\subsection{Multicomponent nucleation}

Now, consider a metastable $N$-component vapor mixture at temperature $T$,  within which liquid droplets  of
an $N$-component solution form as a result of isothermal condensation via nucleation. 
Again, in the framework of the capillarity approximation (whereon  the thermodynamics  of macroscopic theory
of multicomponent condensation is based) the droplets are treated as spheres with sharp
boundaries, internal thermodynamic equilibrium, and with the surface tension of the bulk
liquid of the same composition.$^{7-9}$ 

Let $\nu_i\;\;(i=1,...,N)$ be the number of  molecules of component $i$ in a droplet. Becasue the temperature
of the droplet is constant (and equal to $T$), the state of the droplet is completely determined by the set 
$\{\nu\}\equiv (\nu_1,...,\nu_N)$ which can be thus chosen as the independent variables of state of the
droplet; the capillarity approximation implies that $\nu_i\gg 1\;\;(i=1,...,N)$. The droplet chemical
composition can be characterized by a set $\{\chi\}\equiv (\chi_1,...,\chi_N)$ of mole fractions
$\chi_i\equiv\chi_i(\{\nu\})=\nu_i/\nu\;\;(i=1,...,N)$ (with $\nu=\sum_i\nu_i$ the total number of molecules
in the droplet), of which only $n-1$ are independent because $\sum_i\chi_i=1$.  
The  metastability of the vapor mixture can be characterized by the set of saturation ratios 
$\zeta_i=n_i/n_{i\infty}\;\;(i=1,...,N)$ of its component vapors, where $n_i$ is the partial number density of
molecules of vapor $i$ and $n_{i\infty}$ is the number density of molecules of vapor $i$ 
in equilibrium with its pure bulk liquid at temperature $T$. 

Denote the distribution function of droplets with respect $\{\nu\}$ at time $t$ by  $g(\{\nu\},t)$.   
Introduce the ``complementary" variable 
$\widetilde{\nu_i}$ to include  all but one of the variables of state of a droplet, with  the ``excluded"
variable being $\nu_i$; any function $f$ of variables  $\nu_1,...,\nu_N$ can be thus represented as either
$f(\nu_1,...,\nu_N)$ or $f(\{\nu\})$ or $f(\nu_i,\widetilde{\nu_i})$.    
For instance, 
$g(\{\nu\},t)=g(\nu_i,\widetilde{\nu_i},t)=g(\nu_1,...,\nu_N,t)$.  

If the  droplets exchange matter with the
vapor via absorption and emission of single molecules (as usually assumed in multicomponent CNT),  the
temporal evolution of the distribution $g(\{\nu\},t)$ is governed by the balance equation
\be 
\frac{\partial g(\nu,t)}{\partial t}&=&-\sum_{i=1}^{N}
\left[(W_i^+(\{\nu\})g(\{\nu\},t)-
W^-_i(\nu_i+1,\widetilde{\nu_i})g(\nu_i+1,\widetilde{\nu_i},t))\right.
\nonumber\\
& &\left.- 
(W^+_i(\nu_i-1,\widetilde{\nu_i})g(\nu_i-1,\widetilde{\nu_i},t)-
W_i^-(\{\nu\})g(\{\nu\},t))\right],
\ee
where $W^+_i(\nu)$ and $W^-_i(\nu)\;\;(i=1,..,N)$ are the numbers of molecules of component $i$ that a droplet
$\nu$ absorbs and emits, respectively, per unit time. A differential equation governing the temporal evolution
of $g(\{\nu\},t)$ can be obtained from the discrete balance equation (12) through  the Taylor  series
expansions of $W_i^-(\nu_i\pm 1,\widetilde{\nu}_i),W_i^+(\nu_i\pm 1,\widetilde{\nu}_i)$,  and $g(\nu_i\pm
1,\widetilde{\nu}_i, t)$ (on its RHS) with respect to the deviation of their arguments from 
$\nu_i\;\;(i=1,...,N)$. 

According to classical thermodynamics, the equilibrium distribution function has the form
\beq
g_e(\{\nu\})=n_{\mbox{\tiny f}}\exp[-F(\{\nu\})], 
\eeq
where $n_{\mbox{\tiny f}}$ is the normalization factor and $F(\{\nu\})$ is the free energy of formation of a
droplet $\nu$ (in units of $k_BT$). It can be written in the form$^{1,8,9}$
\beq 
F(\{\nu\})=-\sum b_i\nu_i+a(\{\nu\})(\sum_{i}\nu_i)^{2/3},
\eeq
where $b_i\equiv b_i(\{\chi\})=\ln[\zeta_i/\chi_if_i(\{\chi\})]\;\;(i=1,..,N)$,  $f_i(\{\chi\})$ is the
activity coefficient of component $i$ in the droplet,  and
$a(\{\chi\})=4\pi(3v_l/4\pi)^{2/3}(\sigma(\{\chi\})/k_BT)$, with $v_l(\{\chi\}$ being the
composition-dependent average volume per molecule in the
liquid phase and $\sigma(\{\chi\})$ the droplet surface tension (assumed equal to the surface tension of
a bulk liquid solution of droplet composition $\{\chi\})$. 

The function  $F=F(\nu_1,..,\nu_N)$ determines a free-energy surface in an (N+1)-dimensional  space.  Under
conditions when condensation occurs via    nucleation, it  has the form of a hyperbolic paraboloid
(``saddle-like" shape  in three dimensions).  A droplet, whereof the variables $(\nu_1,...,\nu_N)$  coincide
with the coordinates of the saddle point, is called ``nucleus"; 
these coordinates are determined as the
solution of $N$ simultaneous equations 
\beq 
F'_i(\{\nu\}|_{c}\equiv\left. \frac{\partial F}{\partial \nu_i}\right|_{c}=0\;\;(i=1,..,N).
\eeq  
where $F'_{i}= \partial F(\{\nu\})/\partial \nu_i\;\;(i=1,..,N)$. 
Quantities for the  nucleus will be again marked
with the subscript ``c".

Let us define the quadratic approximation (QA) region $\Omega_{2\nu}$ in the space of variables 
$\{\nu\}$ as the vicinity of  
the saddle point within which 
$F-F_c$ can be accurately approximated 
as a quadratic form 
\beq F-F_c=\frac1{2}\sum_{i,j=1}^NF''_{ijc}\De\nu_i\De\nu_j\;\;\;(\{\nu\}\in\Omega_{2\nu}),
\eeq 
where $F''_{ij}=\partial^2 F/\partial\nu_i \partial\nu_j\;\;(i,j=1,..,N)$   
and  $\De\nu_i\equiv \nu_i-\nu_{ic}\;\;(i=1,..,N)$.
In this approximation, 
the equilibrium distribution can be represented as 
\beq 
g_e(\{\nu\})\simeq g_e(\{\nu_c\})\exp[-\frac1{2}\sum_{i,j=1}^NF''_{ijc}\De\nu_i\De\nu_j] 
\;\;\;\;\;(\{\nu\}\in\Omega_{2\nu}). 
\eeq 

Approximation (16) is equivalent to neglecting the cubic and higher order terms in 
the Taylor series expansion of $F(\{\nu\})$ with respect to deviations $\De\nu_i$ in the vicinity of the
saddle point. 
Therefore, considering that $1/3$ is already much smaller than $1$, the QA region $\Omega_{2\nu}$, 
wherein approximation (16) 
is acceptable, can be determined by the condition 
\beq 
\ep_{\mbox{\tiny 32}}(\{\nu\})\lesssim \frac1{3},
\eeq 
where  
\beq 
\ep_{\mbox{\tiny 32}}(\{\nu\})=\frac{|\sum_{i,j,k=1}^{N}b_{ijk}(\De\nu_i)(\De\nu_j)(\De\nu_k)|}
{|\sum_{i,j=1}^Na_{ij}(\De\nu_i)(\De\nu_j)|}
\eeq
with 
\beq 
a_{ij}\equiv \left. \frac1{2!}\frac{\partial^2 F(\{\nu\})}{\partial \nu_i\partial\nu_j}\right|_c\;,\;\;\;\;
b_{ijk}=\left. \frac1{3!}\frac{\partial^3 F(\{\nu\})}{\partial \nu_i\partial\nu_j\partial\nu_k}\right|_c
\;\;\;(i,j,k=1,..,N) 
\eeq

Let us define the saddle-point (SP) region $\Omega_{\nu}$ in the space of variables $\{\nu\}$ as the minimal 
vicinity  of the saddle point within which the equilibrium distribution $g_e(\{\nu\}$ changes substantially. 
According to eq.(17), its boundary should thus satisfy the constraint (analogous to constraint (4) of  the
unary nucleation theory$^{4}$) 
\beq 
{\bf |\De\nu}^{\mbox{\tiny T}}{\bf A}{\bf \De\nu}|\equiv 
|\frac1{2}\sum_{i,j=1}^NF''_{ijc}(\nu_i-\nu_{ic})(\nu_j-\nu_{jc})|=1, 
\eeq
where the matrix notation was introduced 
with a real symmetric  $N\times N$-matrix ${\bf A}=[a_{ij}]\;\;(i,j=1,..,N)$ 
and a  real column-vector ${\bf \De\nu}=[\De\nu_i]\;\;(i=1,..,N)$ of length N,   
where ${\bf A}=[a_{ij}]\;\;(i,j=1,..,N)$ is 
a real symmetric  $N\times N$-matrix 
and ${\bf \De\nu}=[\De\nu_i]\;\;(i=1,..,N)$ is a  real column-vector of length N,  
the superscript ``$\mbox{T}$" marking  the transpose of a matrix or vector. 

Because the matrix ${\bf A}$ is real and symmetric, it is orthogonally diagonalizable, according to the spectral
theorem.$^{10}$ Therefore, there exists   a real orthogonal  $N\times N$-matrix  ${\bf P}\equiv
[p_{\alpha\gamma}]\;\;(\alpha,\gamma=1,..,N)$  (such that ${\bf P}^{-1}={\bf P}^T$)   diagonalizing the matrix 
${\bf A}$,
so the matrix  $ {\bf D}={\bf P}^{\mbox{\tiny T}}{\bf A}{\bf P} $ is a real diagonal $N\times N$
matrix (hereafter the Greek subscripts $\alpha,\gamma=1,..,N$  do {\em not}
indicate the relation to the chemical components $1,..,N$ in the system). 
By virtue of the spectral theorem,$^{10}$  the columns of the matrix ${\bf P}$ are linearly
independent orthonormal eigenvectors of   ${\bf A}$. The corresponding eigenvalues
$\la_1,..,\la_N$ are the diagonal elements of ${\bf D}$.  When the free energy surface has the shape of a
hyperbolic paraboloid,  
one of these eigenvalues 
is negative (say, $\la_1<0$), while all others are positive, hence $\det({\bf A})<0$. 

Let us introduce the new variables $\{ x\}\equiv (x_1,..,x_N)$ as   
\beq  x_{\alpha}=\sum_{i=1}^Np_{i\alpha}\De\nu_{i}\;\;\;(\alpha=1,..,N),\eeq 
comprising a column-vector ${\bf x}\equiv [x_{\alpha}]\;\;(\alpha=1,..,N)$ of length $N$.  
Because the difference
$F-F_c$ does not depend on the choice of independent variables of state of a droplet, and  ${\bf
\De\nu}^{\mbox{\tiny T}}{\bf A}{\bf \De\nu}={\bf x}^{\mbox{\tiny T}}{\bf D}{\bf x}$,  approximation (16) for
$F$  in variables $\{ x\}$ becomes 
\beq 
F-F_c=\sum_{\alpha}\la_{\alpha} x_{\alpha}^2\;\;\;(\{x\}\in\Omega_{2x}), 
\eeq 
and approximation (17) for the equilibrium distribution transforms into an approximation for the equilibrium
distribution $q_{\mbox{\tiny e}}(\{x\})$ in variables $\{x\}$ (with a new normalization factor $n_x$):   
\beq q_{\mbox{\tiny e}}(\{{\bf x}\})\simeq q_{\mbox{\tiny e}}(\{{\bf x_c}\})
\exp[-\sum_{\alpha}\la_{\alpha}x_{\alpha}^2]\;\;\;(\{x\}\in\Omega_{2x}). 
\eeq  

Thus, the quadratic form in eq.(21), determining the boundary of the SP region $\Omega_{\nu}$ in variables
$\{\nu\}$, becomes  a diagonal quadratic form in variables $\{x\}$. That only one eigenvalue of ${\bf A}$ is
negative ($\la_1<0$), whereas all others are positive ($\la_{\alpha}>0\;\;(\alpha\ne 1)$), allows one to  
identify $x_1$ as the {\em single} thermodynamically unstable variable and $x_2,..,x_N$ as thermodynamically
stable ones. This allows one to impose standard boundary condition on the multidimensional kinetic
equation,$^{11,12}$ 
\be 
\frac{q(\{{\bf x}\},t)}{q_e(\{{\bf x}\})}&=&
\left\{
\begin{array}{ll} 1 & \;\;(x_1\rightarrow -\infty\;\;
\mbox{and}\;\;\sum_{\alpha}\la_{\alpha}x_{\alpha}^2<0),\\
0 & \;\;(x_1\rightarrow \infty\;\;
\mbox{and}\;\;\sum_{\alpha}\la_{\alpha}x_{\alpha}^2<0), 
\end{array} 
\right.\\ 
q(\{{\bf x}\},t)&\rightarrow & 0\;\;\;\; \mbox{as}\;\;\;\; 
\sum_{\alpha}\la_{\alpha} x_{\alpha}^2\rightarrow \infty.\nonumber
\ee  
requiring that in variables $\{x\}$ the distribution function $q(\{x\},t)$ of small sub-critical droplets 
smoothly transition into the equilibrium distribution, whereas for large super-critical droplets $q(\{x\},t)$
smoothly transition into stationary distribution (by convention, sub-critical and super-critical droplets are
assigned negative and positive values, respectively, of the unstable variable $x_1$; in specific applications,
these signs depend on the coefficients of transformation (22), i.e., on the  orthogonal matrix ${\bf P}$).

These boundary conditions to eq.(12) are imposed on the boundary  of the SP region
$\Omega_{\nu}$ of substantial change of $g(\{\nu\},t)$.  
In variables $\{\bf x\}$, the constraint  
\beq  
|\sum_{\alpha=1}^{N}\la_{\alpha}x_{\alpha}^2|=1\;\;\;
(\la_1<0,\;\la_{\alpha}>0\;\;(\alpha\ne 1))\eeq 
will determine the boundary of SP region $\Omega_{x}$ in variables $\{ x\}$; this equation is 
simpler than eq.(21).    Once the boundary of the SP region is determined in variables $\{ x\}$, it can be
also  found in variables $\{\nu\}$ via transformation (22). 

One can then evaluate the accuracy of approximation (16) within the SP region $\Omega_{\nu}$ by calculating
the ratio $\eps_{\mbox{\tiny 32}}(\{\nu\})$ for $\{\nu\}\in \Omega_{\nu}$. According to eq.(18),  the
boundaries of the QA region $\Omega_{2\nu}$, where this approximation is acceptably accurate,  are determined
by the equality $\eps_{32}(\{\nu\})=1/3$. 

Strictly speaking,  approximation (16) is needed in the entire SP region  $\Omega_{\nu}$ (at boundaries of
which conditions (25) are imposed) in order for the kinetic equation of CNT therein to have the
Fokker-Planck form with its force coefficients being linear functions of $\{\nu\}$; in other words, it is
necessary that  $\Omega_{\nu}\subseteq \Omega_{2\nu}$. However, 
even if approximation (16) is not fulfilled in  some (relatively small) part(s) of  $\Omega_{\nu}$,  one can
expect the Fokker-Planck equation with linear force coefficients to be an acceptable approximation for the
kinetic equation in the entire $\Omega_{\nu}$ if the parameter 
\beq
\omega=\frac{\mu[\Omega_{\nu}\setminus (\Omega_{\nu}\cap\Omega_{2\nu})]}{\mu[\Omega_{\nu}]}  
\eeq 
(with $\mu[\Omega]$ denoting the measure of a set (region) $\Omega$) is negligibly small. Clearly, the
smaller $\omega$, the more accurate the kinetic equation of CNT. One can roughly assume that this accuracy is
sufficient if approximation (16) fails only in 10\% of $\Omega_{\nu}$ or less, and require that 
\beq
\omega\lesssim 0.1. 
\eeq
In the case of isothermal binary nucleation ($N=2$), the space of droplet variables is two-dimensional, and
the measure $\mu[\Omega]$ of any region $\Omega$ therein is the surface area of $\Omega$; for isothermal
ternary nucleation ($N=3$),  the space of droplet variables is three-dimensional, and the measure
$\mu[\Omega]$ of a region $\Omega$ therein is the volume of $\Omega$.  


In contrast to the unary nucleation theory, one cannot obtain the operator estimates for the derivatives
$\partial g(\{\nu\},t)/\partial \nu_i$ in the Taylor series expansions of $g(\nu_i\pm 1,\widetilde{\nu}_i,t)$
on the RHS of the balance eq.(12) in a straightforward manner because of the presence of mixed terms
$a_{ij}\De\nu_i\De\nu_j\;\;\;(i,j=1,..,N)$ in the exponential of eq.(17) for $g_e(\{\nu\})$. However, 
the lower limits of the half-widths of the SP region  $\Omega_{x}$ in variables $\{{\bf x}\}$ can
be  estimated to be $\De^x_1\equiv 1/\sqrt{|\lambda_1|}, \De^x_2\equiv 1/\sqrt{\lambda_2},..., \De^x_N\equiv
1/\sqrt{\lambda_N}$ along the axes $x_1,x_2,...,x_N$, respectively, so  
\beq 
\frac1{q(\{x\},t)}\frac{\partial q(\{x\},t)}{\partial x_{\alpha}}\sim 
\frac1{q_s(\{x\})}\frac{\partial q_s(\{x\})}{\partial x_{\alpha}}\sim 
\frac1{q_e(\{x\})}\frac{\partial q_e(\{x\})}{\partial x_{\alpha}}\sim \frac1{\De^x_{\alpha}}
\;\;(\alpha=1,..,N). 
\eeq
Therefore, because  
$$
\frac{\partial g(\{\nu\},t)}{\partial \nu_i}=
\sum_{\alpha=1}^N\frac{\partial J\,g(\{{\bf x}\},t)}{\partial x_{\alpha}}
\frac{\partial x_{\alpha}}{\partial \nu_i},
$$
(where $J$ is the Jacobian of transformation ${\bf \De\nu}={\bf P}{\bf x}$) and $\partial
x_{\alpha}/\partial \nu_i=p_{i\alpha}$, one can obtain estimates 
\beq 
\frac1{g(\{\nu\},t)}\frac{\partial g(\{\nu\},t)}{\partial \nu_i}\sim 
\frac1{g_s(\{\nu\})}\frac{\partial g_s(\{\nu\})}{\partial \nu_i}\sim 
\frac1{g_e(\{\nu\})}\frac{\partial g_e(\{\nu\})}{\partial \nu_i}\lesssim 
\sum_{\alpha=1}^Np_{i\alpha}\sqrt{|\lambda_{\alpha}|}.
\;\;\;(i=1,..,N). 
\eeq

Expanding the procedure of Kuni and Grinin$^{4}$ to multicomponent nucleation, 
performing  the Taylor  series expansions of
$W_i^-(\nu_i\pm 1,\widetilde{\nu}_i),W_i^+(\nu_i\pm 1,\widetilde{\nu}_i)$, 
and $g(\nu_i\pm 1,\widetilde{\nu}_i, t)$ on the RHS of eq.(12),   
and taking into account eq.(30), one can show 
that for the resulting differential equation to 
be accurately approximated by the conventional Fokker-Planck equation of multicomponent CNT 
\beq
\frac{\partial g(\{\nu\},t)}{\partial t}=-\sum_{i=1}^NW^+_{ic}\frac{\partial}{\partial 
\nu_i}\left(-F'_i(\{\nu\})-\frac{\partial}{\partial \nu_i}\right)g(\{\nu\},t)
\eeq
with $F'_i(\{\nu\})\;\;(i=1,..,N)$ being linear superpositions of $\De\nu_i \;\;(i=1,..,N)$  
in the SP region $\Omega_{\nu}$,   
the parameters 
\beq
\frac1{\De^{\nu}_i}\equiv\left|\sum_{\alpha=1}^Np_{i\alpha}\sqrt{|\lambda_{\alpha}|}\right|\;\;\;(i=1,..,N), 
\eeq
must fulfill the strong inequalities 
\beq
\frac1{\De^{\nu}_i}\ll 1 \;\;\;\;\;(i=1,..,N),   
\eeq
in addition to the parameter $\omega$ satisfying constraint (28). 

Thus, the parameters  
$\omega$ and  $1/\De^{\nu}_i\;\;\;(i=1,..,N)$  represent the {\em
small parameters} of the macroscopic theory of multicomponent nucleation. The violation of any one of
constraints (28) or (33) will necessitate going beyond the framework of the conventional 
Fokker-Planck  equation with linear force coefficients usually adopted for the kinetic equation 
in the multicomponent CNT.

If constraint (33) on the parameters $1/\De^{\nu}_i\;\;(i=1,..,N)$ is not satisfied for some $i$, then the
kinetic equation will include contributions of the third and higher order  partial derivatives of the
distribution function $g(\{\nu\},t)$ with respect to $\nu_i$. This constraint can be referred to as the SP
region constraint, because it characterizes how smoothly the distribution function varies in the SP region.  
An elegant method (based on the  combination the Enskog-Chapman method and method of complete separation of
variables) for solving such  a non-Fokker-Planck kinetic equation was developed by Kuni and
Grinin$^{13}$  (see also references 14,15 for its applications). 

On the other hand, if the parameter $\omega$ 
does not satisfy constraint (28),  then the QA region $\Omega_{2\nu}$ of  quadratic approximation (16) for
$F(\{\nu\})$ does not cover a {\em sufficiently large} part of the SP region  $\Omega_{\nu}$   
and it is necessary to retain
the cubic and perhaps even higher order (in  $\De\nu_i\;\;(i=1,..,N)$) terms  in  the Taylor series expansion
for $F(\{\nu\})$. This constraint can be referred to as the QA region constraint, because it characterizes
the extent of the QA region.  As a result, the first derivatives $F'_{i}$ in the kinetic  equation (31) will
not be linear superpositions of deviations $\De\nu_i\;\;(i=1,..,N)$ (they will be  quadratic at least, or
of even higher orders). Hence the force coefficients of equation (31) will no longer be linear functions of 
$\{\nu\}$, i.e., the kinetic equation will differ from the conventional  Fokker-Planck equation of
multicomponent CNT. We are not aware of any work addressing the solution of such a   kinetic equation
in the theory of multicomponent nucleation. 

\section{Numerical evaluations}
As a numerical illustration of the foregoing, 
we have carried out calculations for isothermal condensation in five binary systems:\\ 
(a) 1-butanol (component 1) -- 1-hexanol (component 2); \\ 
(b) water (component 1) -- methanol (component 2); \\ 
(c) water (component 1) -- ethanol (component 2); \\ 
(d) water (component 1) -- 1-propanol (component 2); \\ 
(e) water (component 1) -- 1-butanol (component 2); \\ 
These systems were chosen as representative of the nucleation of droplets of ideal (a) and increasingly
nonideal (b)-(e) binary solutions whose physical and chemical properties,  necessary for the evaluation of
parameters $\omega$ and  $\De^{\nu}_i\;\;\;(i=1,..,N)$ in eqs.(27) and (32), are relatively well known. For
each system, the molecular volumes $v_1$ and $v_2$ of pure liquids  were obtained from the density data of
{\it Lide},$^{16}$ and the mean molecular volume of solution in the droplet was approximated as $v=\chi
v_1+(1-\chi)v_2$, with $\chi=\chi_1$.  All calculations were carried out for the same  system temperature
$T=293.15$ K. Although the saturation ratios $\zeta_1$ and $\zeta_2$ were different in different  systems,
they were always chosen so that the height of the nucleation barrier at the saddle point was in the range from
$30$ to $50$, which would ensure a greater than $1$ cm$^{-3}$s$^{-1}$  nucleation rate (according to binary
CNT$^{1,9,17,18}$).

The  surface tension of 1-butanol(1)--1-hexanol(2) solution (which can be considered  as nearly ideal) was
assumed to  depend on $\chi(=\chi_1)$  as 
$\sigma(\chi)=\chi \sigma_1+(1-\chi)\sigma_2, $  where $\sigma_1$ and $\sigma_2$  are the surface tensions of
pure liquid butanol and pure liquid  hexanol, respectively; $\sigma_1=25.39$ dyn/cm was obtained by linear
interpolation of data from  {\it Lide}$^{16}$ and $\sigma_2=26.20$ dyn/cm was taken  from {\it
Gallant}.$^{19}$ The activity coefficients of both butanol and hexanol in this solution were set equal to
unity (ideal solution approximation).  

For the composition dependence of the surface tension in systems (b)-(e) 
we used the expression 
\beq
\sigma(\chi)=a + b/(d-\chi) + c/(d-\chi)^2
\eeq 
(with $\chi=\chi_1$ and the dimension of $\sigma$ dyn/cm), 
where a set of parameters $a,b,c,d$ for each system. 
These parameters were determined with the help of Mathematica 12.1 by fitting 
expression (34) to appropriate experimental data (of Vazquez {\em et al.}$^{20}$ 
for the systems (b)-(d)) and of Teitelbaum {\em et al.}$^{21}$  for the system (e)):\\  
(b) $a=16.3343, b=8.85203, c=-1.85715\times 10^{-7},d=1.160977$ (water(1)--methanol(2));\\
(c) $a=19.6512, b=3.25232, c=-2.24934\times 10^{-8},d=1.0880297$ (water(1)--ethanol(2));\\
(d) $a=23.4678, b=0.43188, c=-2.53012\times 10^{-10},d=1.02205$ (water(1)--1-propanol(2));\\
(e) $a=24.5474, b=-0.0309657, c=0.00338759,d=1.00811323$ (water(1)--1-butanol(2)). 

The composition dependence of the activity 
coefficients in systems (b)-(e) was described by the van Laar equations  
\beq 
\ln f_1(\chi)=\frac{A_{12}}{(1+\frac{A_{12}\chi}{A_{21}(1-\chi)})^2},\,\,
\ln f_2(\chi)=\frac{A_{21}}{(1+\frac{A_{21}(1-\chi)}
{A_{12}\chi})^2}.
\eeq 
with pairs $A_{12}$ and $A_{21}$ from refs.22, 23:\\ 
(b) $A_{12}=0.5619$ and $A_{21}=0.8041$ (water(1)--methanol(2) solution);\\ 
(c) $A_{12}=0.9227$ and $A_{21}=1.6798$ (water(1)--ethanol(2) solution);\\ 
(d) $A_{12}=1.1572$ and $A_{21}=2.9095$ (water(1)--1-propanol(2) solution);\\ 
(e) $A_{12}=1.0996$ and $A_{21}=4.1760$ (water(1)--1-butanol(2) solution).

Some results of numerical calculations are presented in Figures 1-5. 
Saturation ratios $\zeta_1$ and $\zeta_2$ of vapor mixture components are indicated in the figure captions. 

In each Figure, panel a) shows the SP region $\Omega_{x}$ and the QA region $\Omega_{2x}$ in variables
$\{x\}$, whereas panel b) shows the SP region $\Omega_{\nu}$ and the QA  region $\Omega_{2\nu}$ in
variables $\{\nu\}$; both $\Omega_{2x}$ and $\Omega_{2\nu}$ are shown as grayish areas in these Figures. The
solid curves indicate the borders of SP regions,  whereas the dashed ones indicate the boundaries of QA
regions. In panel a) of each Figure, the thin  dashed lines delineate the rectangular central part 
$\Omega^c_{x}$ of
the SP region $\Omega_{x}$ of  half-widths $\De^x_1$ and $\De^x_2$ which were used in calculating the
parameters  $1/\De^{\nu}_1$  and $\De^{\nu}_2$ according to eq.(32). In panel b) of each Figure,  the
corresponding central part $\Omega^c_{\nu}$ of the SP region $\Omega_{\nu}$ is also shown, 
delineated by thin dashed lines forming a parallelogram. The arrows show the direction of the growth of
droplets at the saddle point. 

As evident from these Figures, in each system the QA region only partially covers the SP
region. Moreover, the QA region does not even cover the central parts of the SP region; approximation
(16) fails to hold even on some segments of its sub-critical and super-critical borders, at which the 
stricter boundary conditions to the kinetic equation (31) are imposed. 

Thus, for all the systems studied, the quadratic approximation (16) for $F(\{\nu\})$ is  accurate {\em not} in
the entire SP region, and it is necessary to retain the cubic and perhaps  even higher order (in 
$\De\nu_i\;\;(i=1,..,N)$) terms  in  the Taylor series expansion for $F(\{\nu\})$. As a result, the first
derivatives $F'_{i}$ in the kinetic  equation (31) will not be linear superpositions of deviations
$\De\nu_i\;\;(i=1,..,N)$ (they will be quadratic or even of higher orders). Hence, the force  coefficients of
equation (31) will no longer be linear functions of  $\{\nu\}$, i.e., the kinetic equation will differ from
the conventional Fokker-Planck equation of multicomponent CNT. Therefore, the conventional expression for the
steady-state binary nucleation rate, obtained on the basis of approximation (16), is not adequate for
comparing theoretical predictions with experimental data  in these systems. 

We have also evaluated the parameters $1/\De^{\nu}_1$, $1/\De^{\nu}_2$, and $\omega$ in all systems (a)-(e). 
For simplicity, $\omega$ was estimated from below by calculating the ratio 
\beq 
\widetilde{\omega}=\left.\frac{\mu[\Omega_{\nu}\setminus 
(\Omega_{\nu}\cap\Omega_{2\nu})]}{\mu[\Omega_{\nu}]}\right|_{\nu\in\Omega^c_{\nu}}  
\eeq
only within the central part $\Omega^c_{\nu}$ of the SP region $\Omega_{\nu}$.  As clear from panels
a) of the Figures, calculating the RHS of eq.(27) in increasingly larger enclosures (rectangles) will result
in increasingly larger results, because beginning from some large enough enclosing enclosure the surface area
of the region $(\Omega_{\nu}\cap\Omega_{2\nu})$ will remain constant while the surface area $\Omega_{\nu}$
will continue to increase asymptotically approaching its limiting value, while the RHS of eq.(27) 
asymptotically approaches $\omega$ from below. 
Therefore, one can
guarantee that  $\omega>\widetilde{\omega}$, and if $\widetilde{\omega}\gtrsim 0.1$, then constraint (28) 
will  certainly {\em not} hold. (Recall that the measures of any region $\Omega$ in variables $\{\nu\}$ and
$\{x\}$ are related as $\mu[\Omega(\{\nu\})]=J\mu[\Omega(\{x\})]$). 

Constraint  (33) on the parameters $1/\De^{\nu}_1$ and  $1/\De^{\nu}_2$ is necessary for neglecting the
terms with the third and higher order derivatives in the Taylor series expansions on the RHS of the balance
equation (12) and thus ensuring the Fokker-Planck form of the kinetic equation.  As clear from the Table,
the smallness of these parameters under metastability conditions that we considered is fulfilled well.
However,   they are sensitive to the saturation ratios $\zeta_1$ and $\zeta_2$, so their smallness at 
given $\zeta_1,\zeta_2$ does not guarantee their smallness at different metastability 
of the vapor mixture. 

On the other hand, constraint (28) on the parameter $\omega$  is necessary in order to ensure that the force
coefficients  of the Fokker-Planck equation  are linear functions of droplet variables in the predominant part
of $\Omega_{\nu}$.  As evident from the Table, under considered metastability conditions this
constraint is not satisfied in any of the systems studied. We note again, however, that $\omega$ is   quite
sensitive to the saturation ratios $\zeta_1$ and $\zeta_2$, so its smallness at a given pair of
$\zeta_1,\zeta_2$ does not guarantee its smallness at different metastability of the vapor mixture. 

Thus, both constraints (28) and (33) must be verified at given $T,\zeta_1,\zeta_2$, and only if they hold, can
one confidently use the conventional CNT expression for the binary nucleation rate $J_s$  for purposes of 
comparing theoretical predictions with experimental data. Otherwise, another, more adequate   theoretical
expression for $J_s$ must be obtained by solving a properly modified kinetic equation 
(which may be of non-Fokker-Planck form).

\renewcommand{\arraystretch}{0.75}
\begin{table} 
\renewcommand{\baselinestretch}{1}
{\em }
\renewcommand{\baselinestretch}{1}
{\bf Table:} 
Small parameters $1/\De^{\nu}_1$, $1/\De^{\nu}_2$, and $\widetilde{\omega}\;(<\omega)$  
of the Fokker-Planck approximation with linear force coefficients in the kinetic equation of binary
nucleation at $T=293.15$ K. 
\vspace{2mm} 
\renewcommand{\baselinestretch}{1}
\begin{tabular}{|c|c|c|c|c|c|c|c|}\hline
\multicolumn{2}{|c|}
{ Binary system}& $\zeta_1$ & $\zeta_2$ & $F_c$ & $1/\De^{\nu}_1$ & $1/\De^{\nu}_2$ & 
$\widetilde{\omega}\;\;(\omega >\widetilde{\omega})$  
\\ \hline
\multicolumn{2}{|c|}{1-butanol(1)--1-hexanol(2)}& $2.88$ & $3.41$ & $40.11$ & $0.19$ & $0.04$ & $0.15$
\vspace{-0.3mm} 
\\
\multicolumn{2}{|c|}{ }& $2.07$ & $3.41$ & $49.96$ & $0.16$ & $0.02$ & $0.12 $\vspace{-0.3mm}   
\\
\multicolumn{2}{|c|}{ }& $2.88$ & $1.95$ & $49.95$ & $0.15$ & $0.08$ & $0.19 $\vspace{-0.3mm}    
\\
\multicolumn{2}{|c|}{ }& $4.30$ & $3.41$ & $30.07$ & $0.23$ & $0.11$ & $0.21 $\vspace{-0.3mm}    
\\
\multicolumn{2}{|c|}{ }& $2.88$ & $6.39$ & $30.21$ & $0.25$ & $0.01$ & $0.15 $\vspace{-0.3mm}    
\\
\hline
\multicolumn{2}{|c|}{water(1)--methanol(2)}& $1.45$ & $0.81$ & $40.14$ & $0.01$ & $0.09$ & $0.51$    
\\ 
\multicolumn{2}{|c|}{ }& $1.29$ & $0.81$ & $50.14$ & $0.07$ & $0.04$ & $0.14$    
\\
\multicolumn{2}{|c|}{ }& $1.45$ & $0.68$ & $50.11$ & $0.07$ & $0.06$ & $0.15 $    
\\
\multicolumn{2}{|c|}{ }& $1.71$ & $0.81$ & $30.21$ & $0.12$ & $0.08$ & $0.19 $    
\\
\multicolumn{2}{|c|}{ }& $1.45$ & $1.01$ & $30.17$ & $0.11$ & $0.05$ & $0.18 $    
\\\hline
\multicolumn{2}{|c|}{water(1)--ethanol(2)}& $1.52$ & $0.9$ & $39.81$ & $0.07$ & $0.08$ & $0.21 $    
\\ 
\multicolumn{2}{|c|}{ }& $1.39$ & $0.90$ & $50.22$ & $0.06$ & $0.05$ & $0.19 $    
\\
\multicolumn{2}{|c|}{ }& $1.52$ & $0.69$ & $49.93$ & $0.05$ & $0.09$ & $0.18 $    
\\
\multicolumn{2}{|c|}{ }& $1.71$ & $0.90$ & $30.14$ & $0.09$ & $0.12$ & $0.24 $    
\\
\multicolumn{2}{|c|}{ }& $1.52$ & $1.21$ & $30.07$ & $0.09$ & $0.06$ & $0.23 $    
\\\hline
\multicolumn{2}{|c|}{water(1)--1-propanol(2)}& $1.3$ & $1.9$ & $39.86$  & $0.01$ & $0.06$ & $0.42 $    
\\ 
\multicolumn{2}{|c|}{ }& $1.1$ & $1.9$ & $49.93$ & $0.07$ & $0.01$ &  $0.14$   
\\
\multicolumn{2}{|c|}{ }& $1.3$ & $1.55$ & $50.82$ & $0.01$ & $0.04$ & $0.54$   
\\
\multicolumn{2}{|c|}{ }& $1.45$ & $1.9$ & $30.35$ & $0.07$ & $0.06$ & $0.53$  
\\
\multicolumn{2}{|c|}{ }& $1.3$ & $2.4$ & $29.89$ & $0.09$ & $0.03$ & $0.30$    
\\\hline
\multicolumn{2}{|c|}{water(1)--1-butanol(2)}& $1.32$ & $2.6$ & $40.07$ & $0.09$ & $0.02$ & $0.28 $    
\\ 
\multicolumn{2}{|c|}{ }& $1.02$ & $2.6$ & $50.21$ & $0.10$ & $0.01$ & $0.13$    
\\
\multicolumn{2}{|c|}{ }& $1.32$ & $2.16$ & $50.03$ & $0.02$ & $0.06$ & $0.32$    
\\
\multicolumn{2}{|c|}{ }& $1.48$ & $2.95$ & $30.45$ & $0.01$ & $0.09$ & $0.74$    
\\
\multicolumn{2}{|c|}{ }& $1.32$ & $3.37$ & $30.25$ & $0.12$ & $0.03$ & $0.27$    
\\
\hline
\end{tabular}
\renewcommand{\baselinestretch}{2}
\end{table}

\section{Concluding remarks}

We have attempted to shed some light on the conditions necessary for the Fokker-Planck equation
with linear force coefficients to be an adequate approximation to the kinetic equation of nucleation in a
macroscopic theory of  isothermal homogeneous multicomponent condensation. Starting with a discrete equation
of balance, governing the temporal evolution of the distribution function of an ensemble of multicomponent
droplets, and reducing it (by means of Taylor series expansions) to the differential form in the vicinity of
the saddle point of the free energy surface,  we have obtained the constraints necessary for the resulting
kinetic equation to have the form of the Fokker-Planck equation and for its force coefficients  to be linear
functions of droplet variables; we have also identified the corresponding ``small" parameters.  

If those (saddle point (SP) region and quadratic approximation (QA) region) constraints are not satisfies,
then either contributions from the third or higher order  partial derivatives of the distribution function in
the kinetic equation will be signinficant (when the SP constraint does not hold) or the force coefficients  of
the Fokker-Planck equation will become non-linear functions of droplet variables (when the QA constraint does
not hold), or both. In any of these cases, the conventional kinetic equation of multicomponent nucleation  and
its predictions would become inaccurate. 

As a numerical illustration, 
we have carried out calculations for isothermal condensation in five binary systems   
(at $T=293.15$ K and vapor mixture metastabilities typical of experimental conditions): 
butanol--hexanol, water--methanol, water--ethanol, water--1-propanol, water--1-butanol.
These systems were chosen as representatives for the nucleation of droplets of ideal (a) and increasingly
nonideal (b-e) binary solutions.  Our results suggest that 
the SP constraint on the smoothness of the droplet distribution in the SP region  is well fulfilled, which
substantiates neglecting the third and higher order derivatives of the distribution function in the
conventional kinetic equation, i.e., its generic Fokker-Planck form. However, the QA constraint on the
quadratic approximation in the Taylor series expansion of the free energy of droplet formation in the saddle
point region is not satisfied; therefore, the force coefficients in that generic Fokker-Planck equation are
not linear functions of droplet variables. Hence, the kinetic equation of binary nucleation does {\em not}
have the form adopted in the binary CNT, so the conventional expressions$^{1-3,8,9,17,18}$ for the
steady-state distribution of binary droplets and steady-state rate of  binary nucleation can be of
insufficient accuracy for making comparisons with experimental data. 

Moreover, numerical calculations show that whether the constraints on the small parameters 
are satisfied or not is quite sensitive to the saturation ratios $\zeta_1,\zeta_2$ and this
sensitivity increases with increasing non-ideality of the liquid solution in droplets. Therefore, for many
conditions of likely interest, it is
necessary to obtain   the steady-state solutions of the modified kinetic equation, going beyond the framework
of the Fokker-Planck equation of CNT due to the non-fulfillment of either the SP region constraint (when the
third or even higher order derivatives of the distribution function are present in the kinetic equation) or
the QA region constraint (when the force coefficients in the generic Fokker-Planck equation are not linear
functions of droplet variables) or both. Clearly, such solutions are needed for consistently 
comparing theoretical predictions and experimental data obtained under conditions when the corresponding
constraints are not fulfilled. This will be an object of our further research.  

\subsubsection*{Acknowledgements}  
{This research project was initiated and partially completed when Eli Ruckenstein, one of the authors, was
still alive but was completed only after he passed away. We thank Buzz I. Dzhikkaity for the  
numerical calculations of the parameter $\widetilde{\omega}$ for all the systems in the Table.} 

\subsubsection*{Data Availability} 
{The data that support the findings of this study are available from the corresponding author upon reasonable
request.} 

\newpage
\subsection*{References} 
\begin{list}{}{\labelwidth 0cm \itemindent-\leftmargin}
\item $1.$ D. Kaschiev,  {\it Nucleation : basic theory with applications}  
(Butterworth Heinemann, Oxford, Boston, 2000). 
\item $2.$ J.W.P. Schmelzer, {\it Nucleation Theory and Applications} (Wiley-VCH Verlag GmbH, 2005)
\item $2a.$ V.V Slezov, {\it Kinetics of First-Order Phase Transitions} 
(Wiley-VCH, Berlin, 2009). 
\item $3.$ E. Ruckenstein and G. Berim, {\it Kinetic theory of nucleation} (CRC, New York, 2016).
\item $4.$ A.P. Grinin and F.M. Kuni, 
Vestnik Leningradskogo universiteta. Seriya Fizika, Khimiya (in Russian)   {\bf 22}, 10 (1982).
\item $5.$ F.M. Kuni, A.P. Grinin, and A.K. Shchekin, 
Physica A, {\bf 252}, 67 (1998).
\item $6.$ V.B. Kurasov, 
Physica A, {\bf 280}, 219 (2000).
\item $7.$ J. Lothe and G.M.J. Pound, in {\it Nucleation}; Zettlemoyer, A. C., Ed. (Marcel-Dekker: New York, 1969).  
\item $8.$ H. Reiss, J. Chem. Phys. {\bf 18}, 840 (1950).
\item $9.$ A.E. Kuchma and A.K. Shchekin, 
J. Chem. Phys. {\bf 150}, 054104 (2019).  
\item $10.$ R. G. Horn and C. R. Johnson, {\it Matrix Analysis} 
(Cambridge University Press, Cambridge, 2013). 
\item $11.$ F. M. Kuni, A. A. Melikhov, T. Yu. Novozhilova, and I. A. Terentev, 
Theor. Math. Phys. {\bf 83}(2), 530-542 (1990).
\item $12.$ F.M. Kuni and A.A. Melikhov,  
Theor. Math. Phys. {\bf 81}(2), 1182-1194 (1989).
\item $13.$ A.P. Grinin and F.M. Kuni,   
Theor. Math. Phys. {\bf 80}, 968 (1989).
\item $14.$ Y. S. Djikaev, F. M. Kuni, and A. P. Grinin, 
J. Aerosol Sci. {\bf  30}, 265-277 (1999).
\item $15.$ Y. S. Djikaev, J. Teichmann, and M. Grmela, 
Physica A {\bf  267}, 322-342 (1999).
\item $16.$ D.R. Lide, Ed. {\em CRC Handbook of Chemistry and Physics},  75th Edition (CRC Press: Boca
Raton, 1994-1995).
\item $17.$ D. Stauffer, 
J. Aerosol Sci. {\bf 7}, 319-333 (1976).
\item $18.$ A.A. Melikhov, V.B. Kurasov, Y.S. Dzhikaev, and F.M. Kuni, 
Sov. Phys. Techn. Phys. {\bf 36}, 14 (1991).
\item $19.$ R.W. Gallant, 
Hydrocarbon Process. {\bf 46}, 133-139 (1967).
\item $20.$ G. Vazquez, E. Alvarez, J.M. Navaza  J. Chem. Eng. Data {\bf 40}(3), 611-614 (1995).
\item $21.$ B.Y. Teitelbaum, T.A. Gortalova, and E.E. Siderova, Zh. Fiz. Khim. {\bf 25}, 911-919 (1951).
\item $22.$ J. Gmehling and U. Onken, 
Vapor-Liquid Equilibrium Data Collection, vol. 1, part 1, Dtsch. Ges. fiir Chem. Apparatewesen,
Chem. Tech. und Biotechnol., (Frankfurt, Germany, 1977).
\item $23.$ R.H. Perry and D.W. Green, Eds. {\em Perry's Chemical Engineers Handbook}. 
(McGraw Hill Companies, 1999).

\end{list}

\newpage 
\subsection*{Captions} 
to  Figures 1 to 5 of the manuscript 
{\sc
``On the Fokker-Planck approximation in the kinetic equation of multicomponent classical nucleation theory"
}  
by {\bf  Y. S. Djikaev}, {\bf E. Ruckenstein}, and {\bf Mark Swihart}
\subsubsection*{}
\vspace{-1.0cm}   
Figure 1. 
The saddle point (SP) and quadratic approximation (QA) regions of the space of droplet variables for binary
nucleation in 1-butanol(1)--1-hexanol(2) vapor mixture at $T=293.15$ K, $\zeta_1=2.88$, and $\zeta_2=3.49$.  
a)  The SP region $\Omega_{x}$ and the QA region $\Omega_{2x}$ in variables $\{x\}$.  b) The SP region
$\Omega_{\nu}$ and the QA  region $\Omega_{2\nu}$ in variables $\{\nu\}$.  The dashed and solid curves 
indicate the boundaries of the SP and QA regions, respectively. Both $\Omega_{2x}$ and $\Omega_{2\nu}$ are
shown as grayish areas. The thin-dashed quadrilateral delineates the central part of the SP region (see the
text); vertices of the same color correspond to  one same droplet in variables $\{x\}$ (a) and $\{\nu\}$ (b).
The arrow points from  the saddle point (red dot) toward super-critical droplets. 
\vspace{0.3cm}\\ 
Figure 2. 
The saddle point (SP) and quadratic approximation (QA) regions of the space of droplet variables for binary
nucleation in water(1)--methanol(2) vapor mixture at $T=293.15$ K, $\zeta_1=1.45$, and $\zeta_2=0.81$.   a) 
The SP region $\Omega_{x}$ and the QA region $\Omega_{2x}$ in variables $\{x\}$.  b) The SP region
$\Omega_{\nu}$ and the QA  region $\Omega_{2\nu}$ in variables $\{\nu\}$. The dashed and solid curves 
indicate the boundaries of the SP and QA regions, respectively. Both $\Omega_{2x}$ and $\Omega_{2\nu}$ are
shown as grayish areas. The thin-dashed quadrilateral delineates the central part of the SP region (see the
text); vertices of the same color correspond to  one same droplet in variables $\{x\}$ (a) and $\{\nu\}$ (b).
The arrow points from  the saddle point (red dot) toward super-critical droplets.  
\vspace{0.3cm}\\ 
Figure 3.  
The saddle point (SP) and quadratic approximation (QA) regions of the space of droplet variables for binary
nucleation in water(1)--ethanol(2) vapor mixture at  $T=293.15$ K, $\zeta_1=1.52$, and $\zeta_2=0.90$.   a) 
The SP region $\Omega_{x}$ and the QA region $\Omega_{2x}$ in variables $\{x\}$.  b) The SP region
$\Omega_{\nu}$ and the QA  region $\Omega_{2\nu}$ in variables $\{\nu\}$. The dashed and solid curves 
indicate the boundaries of the SP and QA regions, respectively. Both $\Omega_{2x}$ and $\Omega_{2\nu}$ are
shown as grayish areas. The thin-dashed quadrilateral delineates the central part of the SP region (see the
text); vertices of the same color correspond to  one same droplet in variables $\{x\}$ (a) and $\{\nu\}$ (b).
The arrow points from  the saddle point (red dot) toward super-critical droplets.
\vspace{0.3cm}\\ 
Figure 4. 
The saddle point (SP) and quadratic approximation (QA) regions of the space of droplet  variables for binary
nucleation in water(1)--1-propanol(2) vapor mixture at $T=293.15$ K,  $\zeta_1=1.30$, and $\zeta_2=1.90$.  a) 
The SP region $\Omega_{x}$ and the QA region $\Omega_{2x}$ in variables $\{x\}$.  b) The SP region
$\Omega_{\nu}$ and the QA  region $\Omega_{2\nu}$ in variables $\{\nu\}$. The dashed and solid curves 
indicate the boundaries of the SP and QA regions, respectively. Both $\Omega_{2x}$ and $\Omega_{2\nu}$ are
shown as grayish areas. The thin-dashed quadrilateral delineates the central part of the SP region (see the
text); vertices of the same color correspond to  one same droplet in variables $\{x\}$ (a) and $\{\nu\}$ (b).
The arrow points from  the saddle point (red dot) toward super-critical droplets.
\vspace{0.3cm}\\ 
Figure 5.  
The saddle point (SP) and quadratic approximation (QA) regions of the space of droplet variables for binary
nucleation in water(1)--1-butanol(2) vapor mixture at $T=293.15$ K, $\zeta_1=1.32$, and $\zeta_2=2.6$.   a) 
The SP region $\Omega_{x}$ and the QA region $\Omega_{2x}$ in variables $\{x\}$.  b) The SP region
$\Omega_{\nu}$ and the QA  region $\Omega_{2\nu}$ in variables $\{\nu\}$. The dashed and solid curves 
indicate the boundaries of the SP and QA regions, respectively. Both $\Omega_{2x}$ and $\Omega_{2\nu}$ are
shown as grayish areas. The thin-dashed quadrilateral delineates the central part of the SP region (see the
text); vertices of the same color correspond to  one same droplet in variables $\{x\}$ (a) and $\{\nu\}$ (b).
The arrow points from  the saddle point (red dot) toward super-critical droplets.

\newpage
\begin{figure}[htp]\vspace{-1cm}
	      \begin{center}
$$
\begin{array}{c@{\hspace{0.3cm}}c} 
              \leavevmode
      	      \vspace{0.cm}
	\leavevmode\hbox{a) \vspace{1cm}} &   
\includegraphics[width=8.3cm]{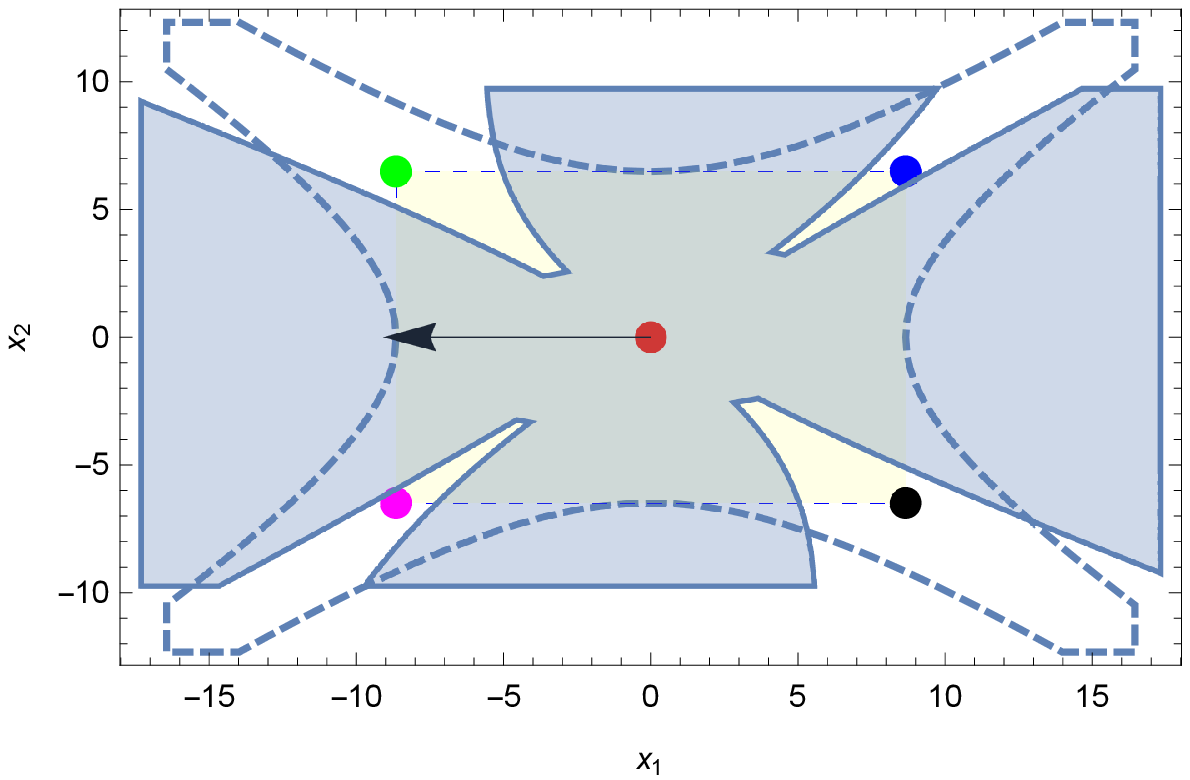}\\ [1.cm] 
      	      \vspace{0.cm}
	\leavevmode\hbox{b) \vspace{1cm}} &  
      	      \vspace{0.0cm}
\includegraphics[width=8.3cm]{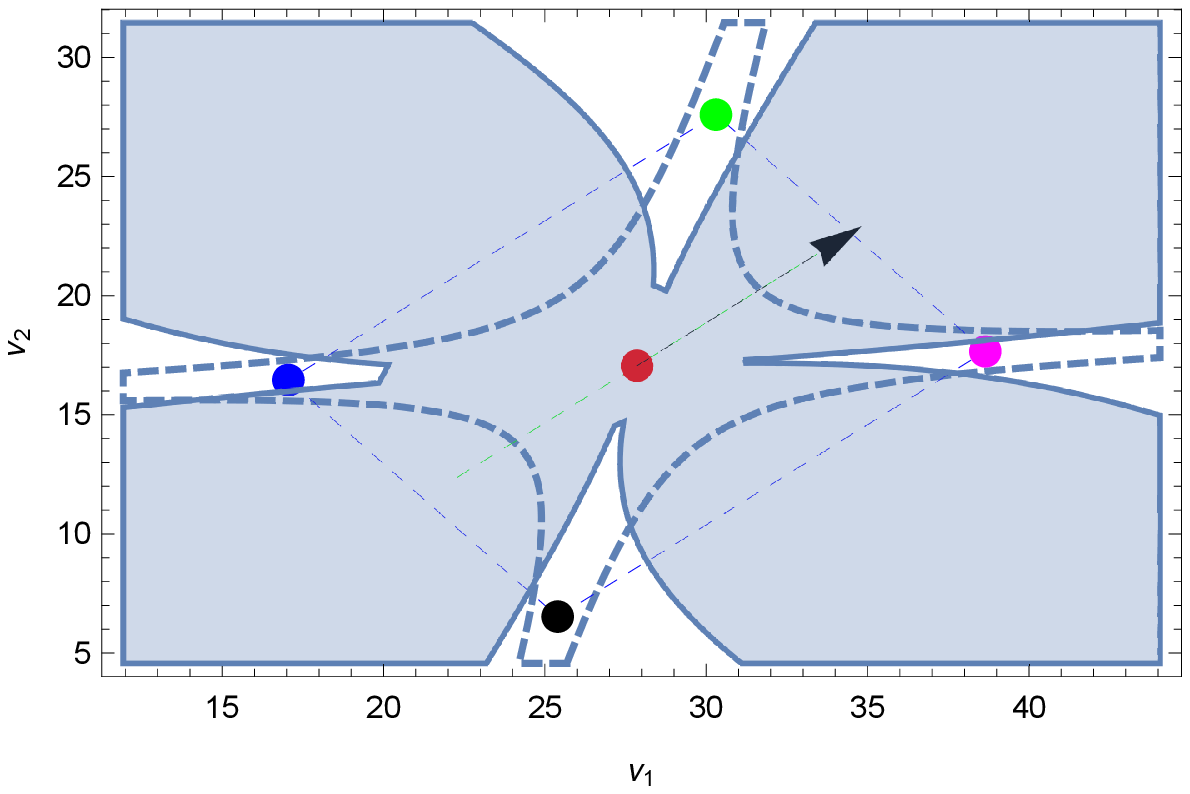}\\ [0.3cm] 
\end{array}  
$$  
	      \end{center} 
            \caption{\small } 
\end{figure}

\newpage
\begin{figure}[htp]\vspace{-1cm}
	      \begin{center}
$$
\begin{array}{c@{\hspace{0.3cm}}c} 
              \leavevmode
      	      \vspace{-0.8cm}
	\leavevmode\hbox{a) \vspace{3cm}} &   
\includegraphics[width=8.7cm]{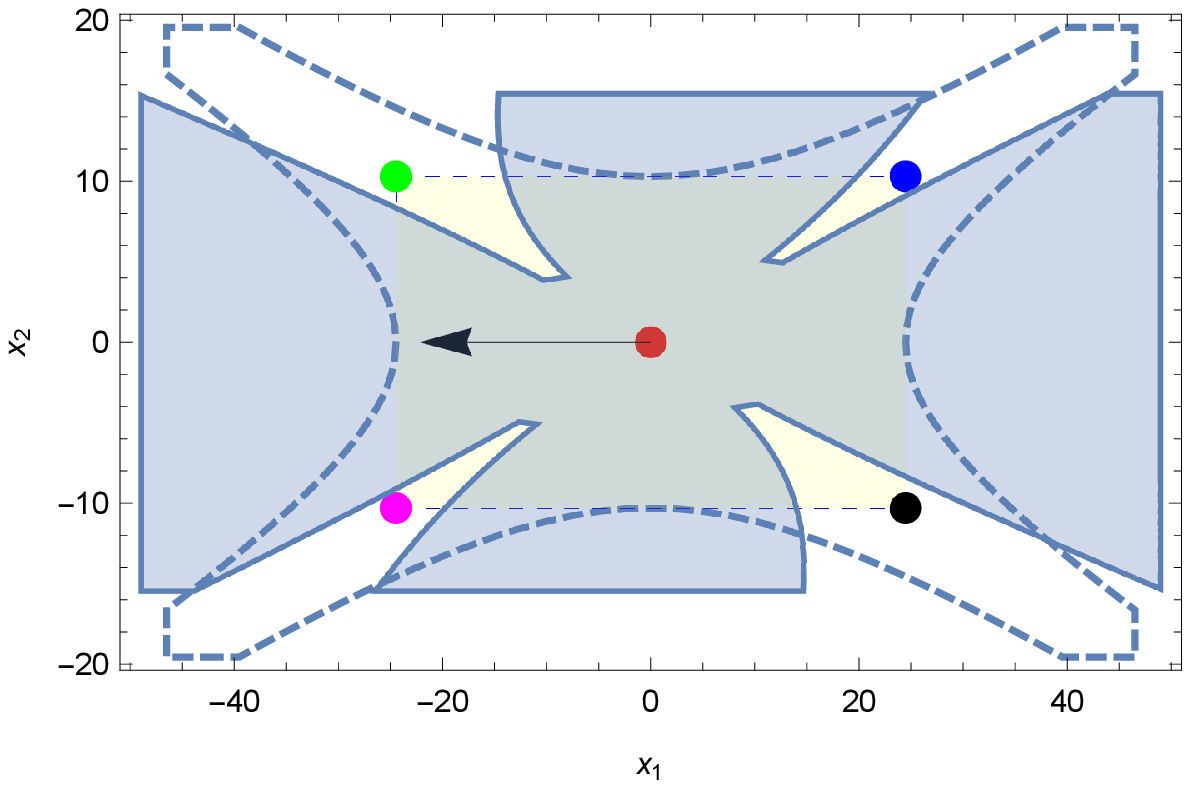}\\ [1.9cm] 
      	      \vspace{0cm}
	\leavevmode\hbox{b) \vspace{3cm}} &  
      	      \vspace{0.0cm}
\includegraphics[width=8.7cm]{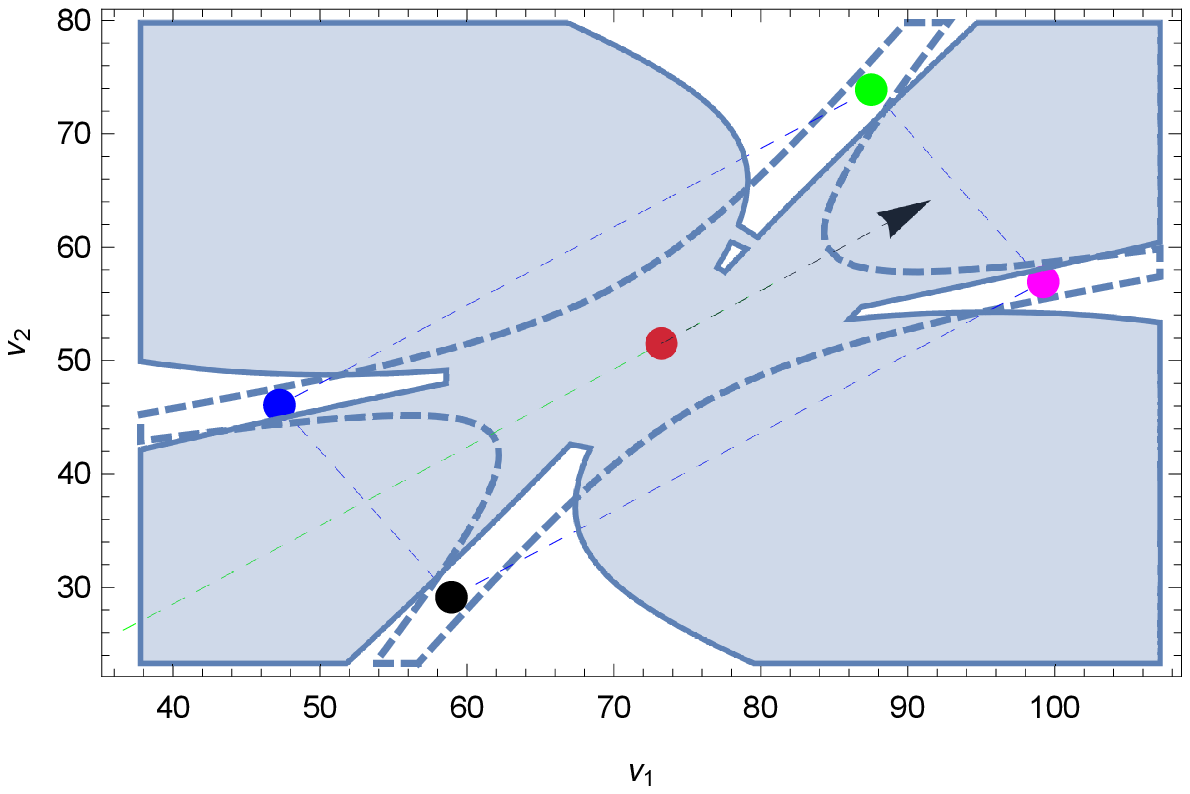}\\ [0.7cm] 
\end{array}  
$$  

	      \end{center} 
            \caption{\small } 

\end{figure}

\newpage
\begin{figure}[htp]\vspace{-1cm}
	      \begin{center}
$$
\begin{array}{c@{\hspace{0.3cm}}c} 
              \leavevmode
      	      \vspace{-0.8cm}
	\leavevmode\hbox{a) \vspace{3cm}} &   
\includegraphics[width=8.7cm]{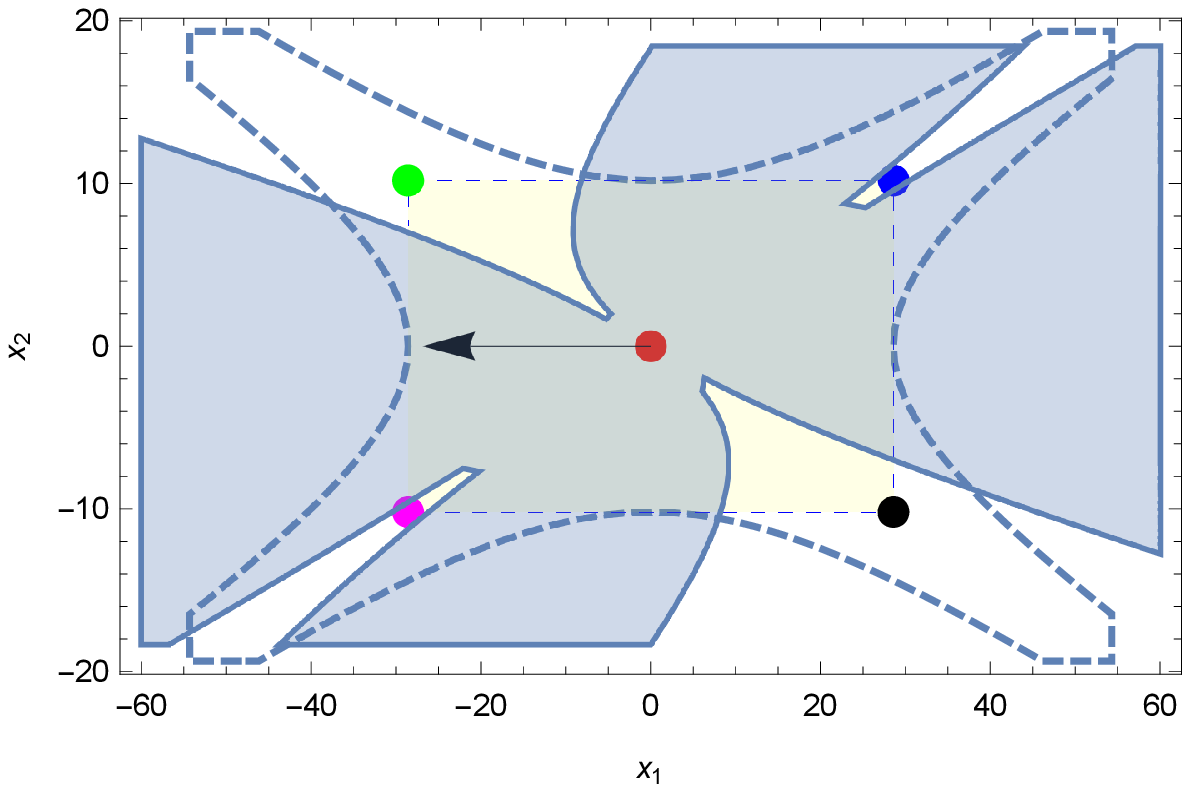}\\ [1.9cm] 
      	      \vspace{0cm}
	\leavevmode\hbox{b) \vspace{3cm}} &  
      	      \vspace{0.0cm}
\includegraphics[width=8.7cm]{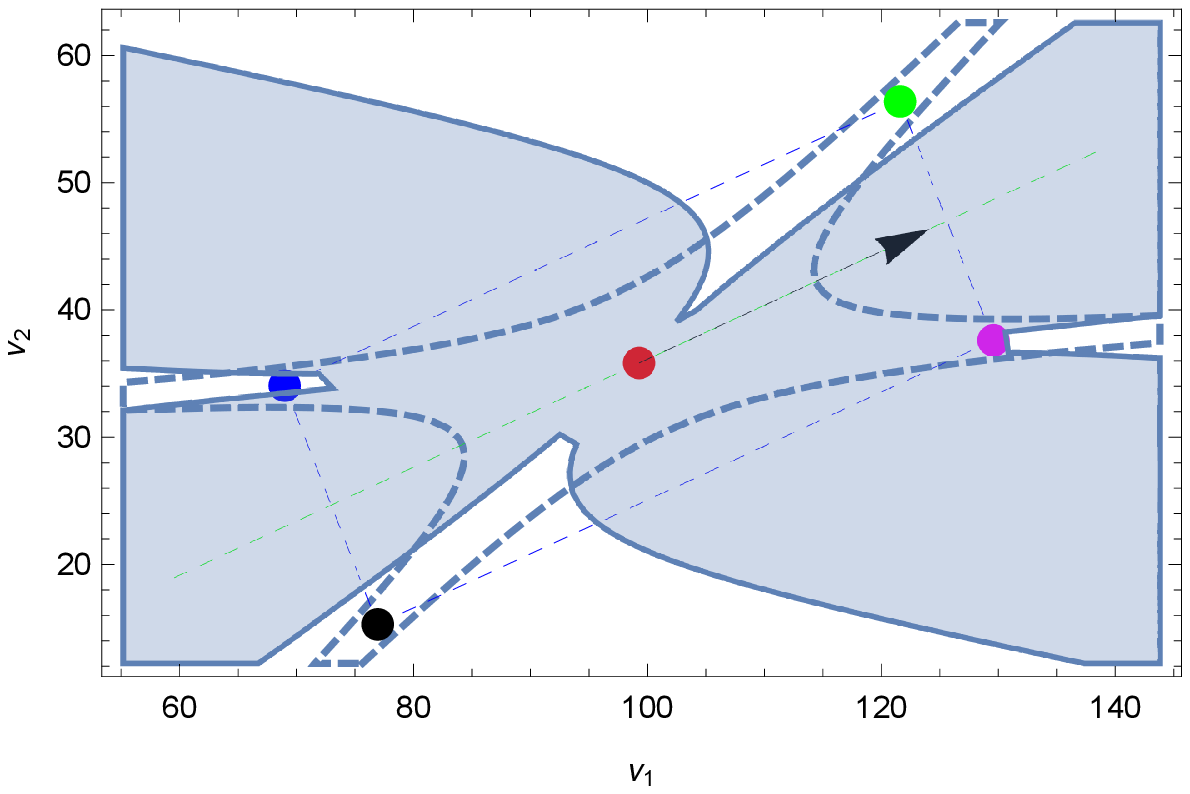}\\ [0.7cm] 
\end{array}  
$$  

	      \end{center} 
            \caption{\small } 

\end{figure}

\newpage
\begin{figure}[htp]\vspace{-1cm}
	      \begin{center}
$$
\begin{array}{c@{\hspace{0.3cm}}c} 
              \leavevmode
      	      \vspace{-0.8cm}
	\leavevmode\hbox{a) \vspace{3cm}} &   
\includegraphics[width=8.7cm]{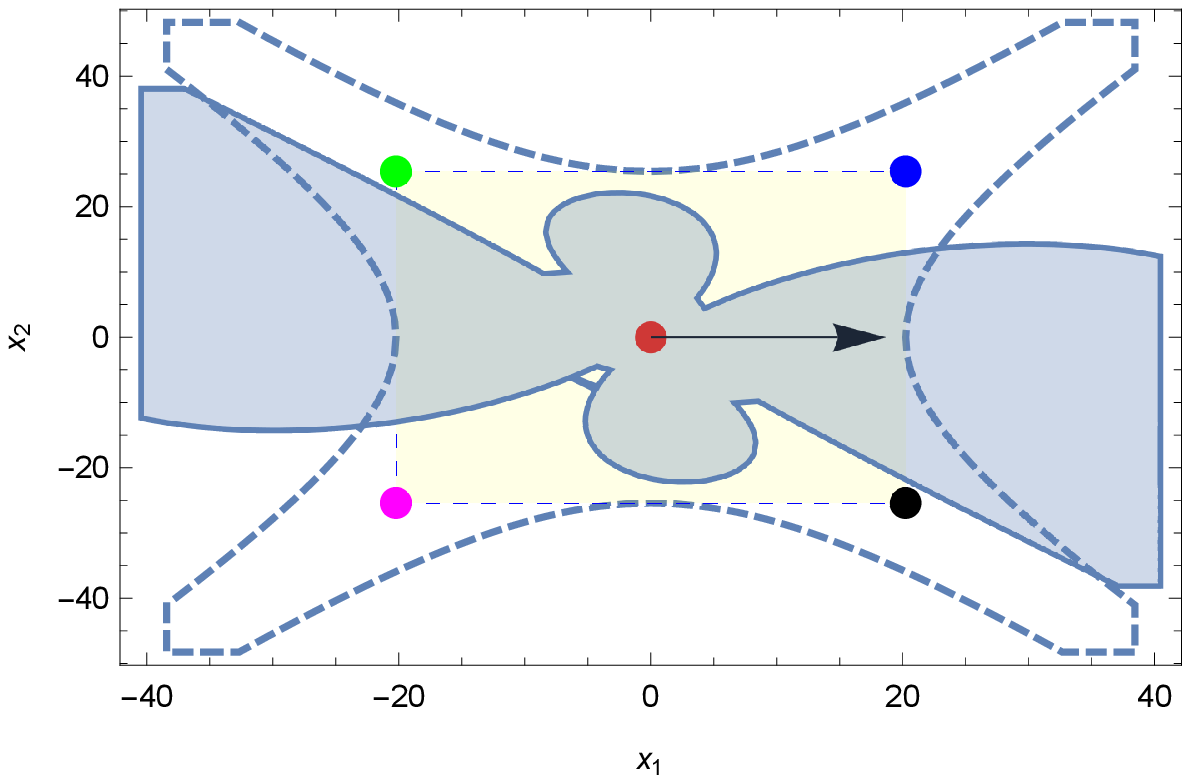}\\ [1.9cm] 
      	      \vspace{0cm}
	\leavevmode\hbox{b) \vspace{3cm}} &  
      	      \vspace{0.0cm}
\includegraphics[width=8.7cm]{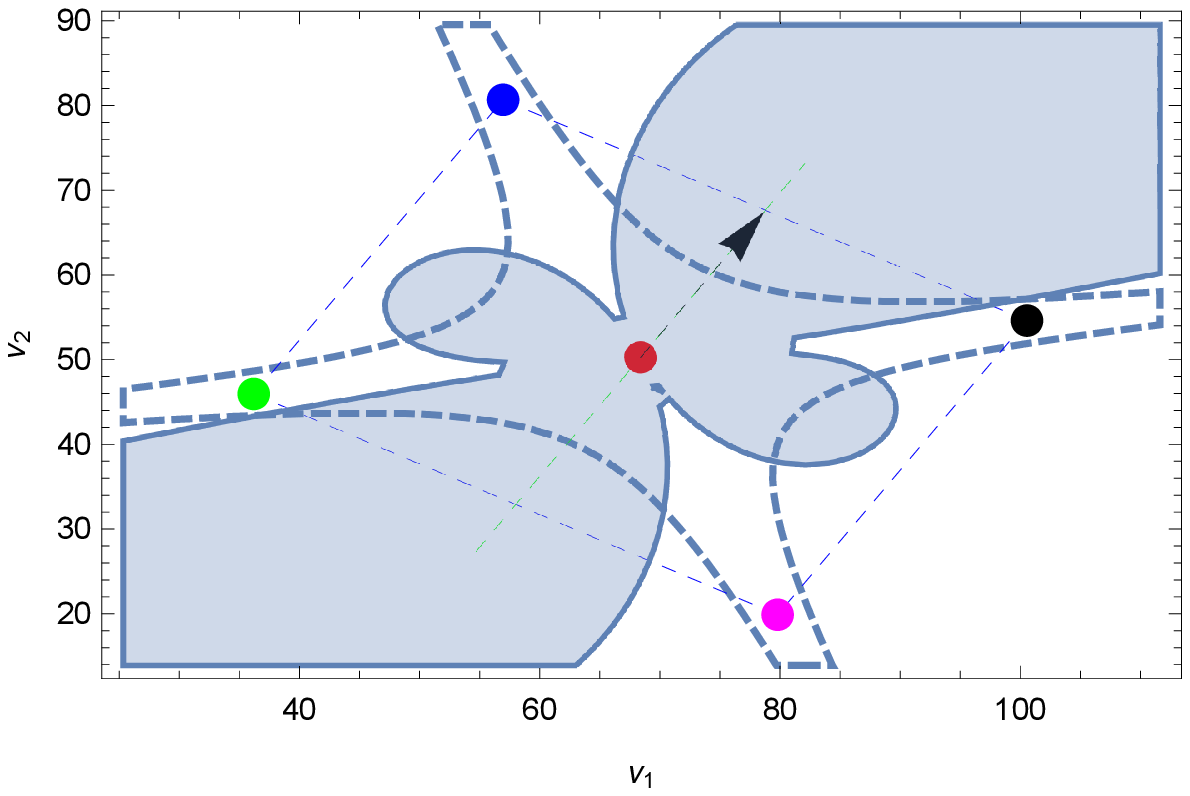}\\ [0.7cm] 
\end{array}  
$$  

	      \end{center} 
            \caption{\small } 

\end{figure}

\newpage
\begin{figure}[htp]\vspace{-1cm}
	      \begin{center}
$$
\begin{array}{c@{\hspace{0.3cm}}c} 
              \leavevmode
      	      \vspace{-0.8cm}
	\leavevmode\hbox{a) \vspace{3cm}} &   
\includegraphics[width=8.7cm]{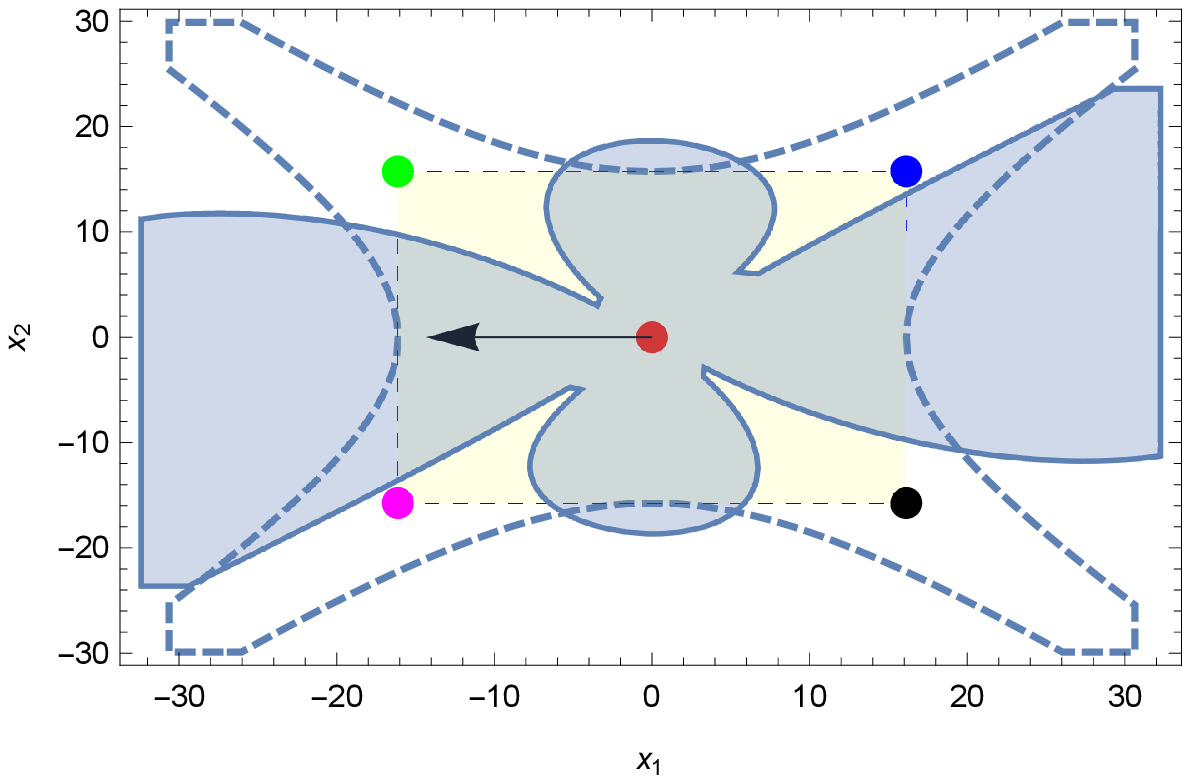}\\ [1.9cm] 
      	      \vspace{0cm}
	\leavevmode\hbox{b) \vspace{3cm}} &  
      	      \vspace{0.0cm}
\includegraphics[width=8.7cm]{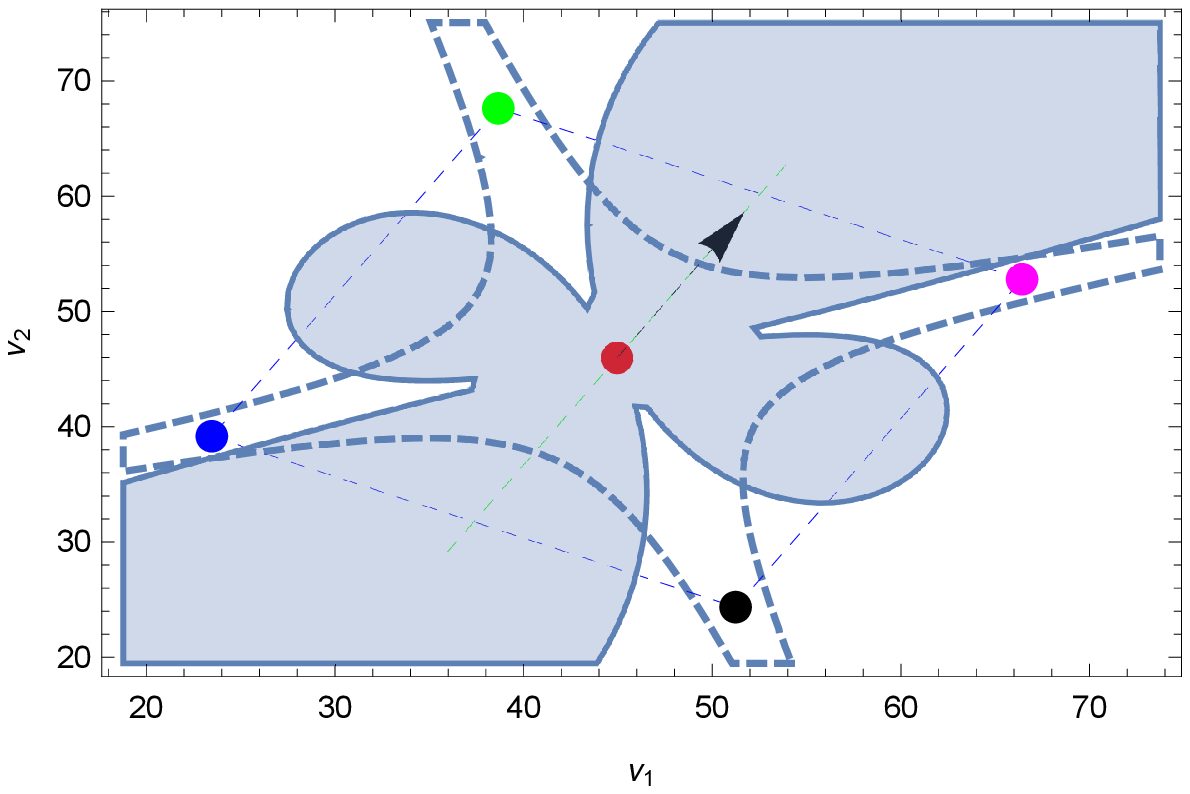}\\ [0.7cm] 
\end{array}  
$$  

	      \end{center} 
            \caption{\small } 

\end{figure}

\end{document}